\documentclass{ijuc}
\usepackage[pdftex]{graphics}

\begin{document}

\title{Classical Simulation of Quantum Adiabatic Algorithms using Mathematica on GPUs }

\author{Sandra D\'iaz-Pier\inst{1}\email{a00455576@itesm.mx}
\and Salvador E. Venegas-Andraca\inst{1}\email{sva@mindsofmexico.org} 
\and  Jos\'e Luis G\'omez-Mu\~noz\inst{1}\email{jose.luis.gomez@itesm.mx}
}

\institute{Quantum Information Processing Group, Tecnol\'{o}gico de Monterrey Campus Estado de M\'{e}xico.
Carretera Lago de Guadalupe Km 3.5, Atizap\'{a}n de Zaragoza, Estado de M\'{e}xico, M\'{e}xico.
}

\def\received{Received 5 March 2011; In final form } 

\maketitle

\begin{abstract}

In this paper we present a simulation environment enhanced with parallel processing which can be used on personal  computers, based on a high-level user interface developed on  Mathematica{\small$^{\copyright}$} which is connected to C++ code in order to make our platform capable of communicating with a Graphics Processing Unit. We introduce the reader to the behavior of our proposal by simulating a quantum adiabatic algorithm designed for solving hard instances of the 3-SAT problem. We show that our simulator is capable of significantly increasing the number of qubits that can be simulated using classical hardware. Finally, we present a review of currently available classical simulators of quantum systems together with some justifications, based on our willingness to further understand processing properties of Nature, for devoting resources to building more powerful simulators.

\end{abstract}

\keywords{classical simulation of quantum algorithms, adiabatic quantum computation, GPU, Mathematica, symbolic computation, natural and artificial parallel processes.}

\section{Introduction}

Quantum Computation, one of the most recent joint ventures between physics and computer science, is a promising emerging branch of science and technology aiming at providing us with algorithms and experimental devices that allow us to exploit quantum effects of physical systems, in order to perform simulations and calculations. Quantum Computing promises great advances in the solution of some problems for which we know no efficient algorithms under the classical computer models and systems we currently have \cite{nielsen00}. Moreover, current results and developments on both theoretical (e.g.  \cite{nielsen00,gruzka00,kitaev02,lanzagorta09,childs10,jordan05}) and experimental (e.g.  \cite{joo06,lanyon07,prevedel07,branderhorst08,porras06,benhelm08,weber10,zahringer10}) arenas of quantum computing have resulted in an increased interest of several applied scientific communities to cross-fertilize their own fields with techniques and ideas from this discipline (e.g. \cite{trugenberger01,trugenberger02a,trugenberger02b,arndt09,cooper11}.)

One of the main problems a computer scientist faces when learning and working on the development of quantum algorithms is the counterintuitive behavior of quantum mechanical systems. For this reason, together with the need to test experimental proposals before implementing them, building powerful classical computer platforms for the simulation of quantum systems is crucial in order to develop intuition about the behavior of quantum systems used for computational purposes, as well as to realize the approximate behavior of practical implementations of quantum algorithms. Particularly, quantifying resources required to process information and/or to compute a solution, i.e. to assess the complexity of the process, is a prioritized research area, as it allows us to estimate implementation costs, as well as to compare problems by comparing the complexity of their solutions. In summary, building simulators for quantum algorithms in classical computers would allow the scientific community to study and analyze the expected behavior and potential of these algorithms on future quantum computers. 

Developing classical computer simulations of quantum algorithms usually has two drawbacks: i) running such simulations of quantum algorithms is frequently a highly demanding task (i.e. an exponential amount of computational resources is typically needed for exact simulations), and ii) due to the computer languages typically used for such classical simulations (e.g. C, C++, Phyton), computer scientists usually have a hard time focusing on solving the problem in mind because of the overwhelming low-level programming details, i.e. high-level languages and better interfaces are needed.

In this paper we introduce a parallel hardware and software simulation platform for quantum algorithms with high-level user interfaces.  Our simulation environment is based on a high-level user interface developed on  Mathematica{\small$^{\copyright}$} which is connected to C++ code in order to make our platform capable of communicating with a Graphics Processing Unit (GPU.)

Our simulation environment is designed to take full advantage of multi-core parallel processing capabilities on the GPU in order to enhance the performance of such classical simulations thus giving scientists the option to work with more extensive problems in less time and without the need to access grid or cloud infrastructures. The high level interface, compiled as a Mathematica{\small$^{\copyright}$} add-on called Quantum{\small$^{\copyright}$}, allows scientists to express their algorithms using the Dirac notation without having to translate them into a matrix form. Then, we use Mathlink{\small$^{\copyright}$} to send the information from Mathematica to C++ code prepared to deploy the parallel tasks to the multiple cores contained in the GPU. Our CUDA interface allows users to communicate with kernels prepared to solve specific problems or with the linear algebra CUDA libraries CUBLAS and CULA. 

Our proposal could be used by quantum scientists to enhance the performance of quantum computing simulations using a single PC equipped with an NVIDIA{\small$^{\copyright}$} CUDA-compatible GPU. Moreover, our very user-friendly interfaces hide the technical, i.e. coding complexity, details of building parallel algorithms for GPUs by creating kernel. In the example we present on this paper,  we have designed such kernels to simulate hard instances of an NP-complete problem, 3-satisfiability problem (3-SAT) \cite{garey79,sipser05}. We present results for benchmarks performed with a variety of instances of the 3SAT problem running with our simulator on the CPU and the GPU.

The rest of this paper is divided as follows: we start by providing the reader with preliminary information about quantum adiabatic computation, the 3-SAT problem and a concise review of classical simulation of quantum algorithms, as these three topics are needed in order to properly describe the structure of our contribution. This section is followed by a reflection on the relationship between natural parallel processing and computer parallel processing, being our comments of this section a contribution towards realizing how massive distributed-parallel computer systems can be used to learn more about Nature and her processes. We then proceed to introduce the reader to the theoretical and practical foundations of our proposal, followed by numerical results produced by simulating a quantum adiabatic algorithm designed to simulate hard instances of 3-SAT. We finish this paper by delivering a conclusions section.

\section{Preliminaries: Quantum Adiabatic Computation, the 3-SAT problem, and a concise review and justification of classical simulation of quantum algorithms}

The purpose of this section is to provide the reader with the preliminary concepts upon which we have built our proposal. We start by delivering the basics of Adiabatic Quantum Computation as that is the universal model of quantum computation we have employed in order to build an example to show the capacities of our simulation platform. Furthermore, we have also used the Hamiltonian proposed in \cite{perdomo11} for solving hard instances of the 3-SAT problem by adiabatic evolution, that is why we also introduce the definition, main characteristics and an example of the 3-SAT problem. We finish this section by providing a concise review of currently existing classical simulators of quantum algorithms.

\subsection{Quantum Adiabatic Computation}
The realization of a robust quantum computer must fulfill several requirements \cite{divincenzo00}, including the development of universal models of quantum computation. Among such models we find Adiabatic Quantum Computation (AQC) \cite{farhi00,farhi01}, a promising paradigm of quantum computing due to its robustness \cite{lidar08,dickson11}, its encouraging results in the study of NP-complete problems  \cite{farhi00,farhi01,hogg03,young08,perdomo11}, and its implementation for the study of statistical mechanical complex problems such as protein folding \cite{perdomo08}.

The goal of AQC algorithms is to transform an initial ground state $|\psi(0)\rangle$ into a final ground state $|\psi(\tau)\rangle$, which encodes the answer to the problem. This is achieved by evolving the corresponding physical system according to the Schr\"{o}dinger equation with a time-dependent Hamiltonian $\hat H(t)$. The AQC algorithm relies on the quantum adiabatic theorem \cite{Messiah,farhi01}, which states that the time propagation of the quantum state will remain very close to the instantaneous ground state $|\psi_{g} (t)\rangle$ for all $t \in [0, \tau]$, whenever $\hat H(t)$
varies \emph{slowly enough} throughout the propagation time $t \in [0, \tau]$ and assuming the ground state manifold does not cross the energy levels which lead to excited states of the final Hamiltonian. Here, we denote by ground state manifold the first $m$ curves associated with the lowest eigenvalue of the time-dependent Hamiltonian for $t \in [0, \tau]$, where $m$ is the degeneracy of the final Hamiltonian ground state. 

Conventionally, the adiabatic evolution path is the linear sweep of $s \in [0,1]$, where $s = t/\tau$:
\begin{equation} \label{eq:conventional-aqc}
\hat{H}(s) = (1-s) \hat{H}_{i} + s \hat{H}_{f}.
\end{equation}
$\hat{H}_{i}$ is usually chosen such that
its ground state is a uniform superposition of all possible $2^{n}$ computational basis vectors. Here, we choose the spin states \{$|q_{i} = 0\rangle,
|{q = 1}\rangle$\}, which are the eigenvectors of $\hat{\sigma}_{i}^{z}$ with eigenvalues +1 and -1, respectively, as the basis vectors. Then the initial ground state
is $|\psi_{g} (0)\rangle = \frac{1}{\sqrt{2^{n}}} \sum_{q_{i} \in \{0,1\}} |q_{n}\rangle |q_{n-1}\rangle \cdots |q_2\rangle |q_1\rangle$. Such an
initial ground state is usually assumed to be easy to prepare and it results in a quantum state with equal probability of all possible solutions.

\subsection{3-SAT}

For $K\geq 3$, K-SAT is an NP-complete problem \cite{garey79,sipser05} and instances of this problem are particularly difficult to solve when the ratio of number of clauses to number of variables is about 4.2  \cite{achlioptas}. Studying the properties of 3-SAT is an important area of research, not only because a polynomial-time solution to 3-SAT would imply {\bf P = NP}, but also because 3-SAT may be used to model problems and procedures in theoretical computer science \cite{acharyya2007} as well as in several areas of applied computer science and engineering like artificial intelligence \cite{gent99,mezard02}. we now provide the reader with a concise introduction of the K-SAT problem together with an example of 3-SAT instances.
\\\\
{\bf The K-SAT Problem}. Let $A=\{e_1, e_2, \ldots, e_n, \bar{e}_1, \bar{e}_2, \ldots, \bar{e}_n \}$
be a set of Boolean variables $E=\{e_i \}$ and their negations $\bar{E}=\{\bar{e}_i \}$. Let us now construct a logical proposition $P$,
defined as $P = \bigwedge_i [(\bigvee_{j=1}^k a_j)] = \bigwedge_i C_i$, where $a_j \in A$, i.e. P is a conjunction of clauses $C_i$ over the set $A$, where each clause consists
of the disjunction of $k$ literals.
Proposition $P$ is a K-SAT instance and the solution of the K-SAT problem, for instance $P$, consists
of finding a set of values for those binary variables upon which $P$ has been built (i.e. a bitstring), so that
replacement of such binary variables for their corresponding binary values makes $P=1$, namely, proposition
$P$ is satisfied.

In order to provide a concise example of how a 3-SAT instance looks like, together with an attempt to show how difficult solving 3-SAT hard instances is, 
let $E=\{x_1, x_2, x_3, x_4, x_5, x_6 \}$ be a set of binary variables and 
consider a 3-SAT instance specified by

$$
\begin{array}{lll}
P &=&  (\bar{x_1} \vee \bar{x_4} \vee \bar{x_5})  \wedge  (\bar{x_2} \vee \bar{x_3} \vee \bar{x_4}) \wedge (x_1 \vee x_2 \vee \bar{x_5})              \wedge  (x_3 \vee x_4 \vee x_5)                   \wedge \\
  & &  (x_4 \vee x_5 \vee \bar{x_6})              \wedge  (\bar{x_1} \vee \bar{x_3} \vee \bar{x_5}) \wedge (x_1 \vee \bar{x_2} \vee \bar{x_5})        \wedge  (x_2 \vee \bar{x_3} \vee \bar{x_6})       \wedge \\
  & &  (\bar{x_1} \vee \bar{x_2} \vee \bar{x_6})  \wedge  (x_3 \vee \bar{x_5} \vee \bar{x_6})       \wedge (\bar{x_1} \vee \bar{x_2} \vee \bar{x_4})  \wedge  (x_2 \vee x_3 \vee \bar{x_4})             \wedge \\
  & &  (x_2 \vee x_5 \vee \bar{x_6})              \wedge  (x_2 \vee \bar{x_3} \vee \bar{x_5})       \wedge (\bar{x_2} \vee \bar{x_3} \vee \bar{x_4})  \wedge  (x_2 \vee x_3 \vee x_6)                   \wedge \\
  & &  (\bar{x_1} \vee \bar{x_2} \vee \bar{x_3})  \wedge  (\bar{x_1} \vee \bar{x_4} \vee \bar{x_5}) \wedge (\bar{x_3} \vee \bar{x_4} \vee \bar{x_6})  \wedge  (\bar{x_4} \vee \bar{x_5} \vee x_6)       \wedge \\
  & &  (\bar{x_2} \vee x_3 \vee \bar{x_6})        \wedge  (x_2 \vee x_5 \vee x_6)                   \wedge (x_3 \vee x_5 \vee \bar{x_6})              \wedge  (\bar{x_1} \vee x_3 \vee \bar{x_6})       \wedge \\
  & &  (x_3 \vee \bar{x_5} \vee x_6)              \wedge  (x_4 \vee x_5 \vee x_6)                   \wedge (x_1 \vee x_2 \vee \bar{x_3}) \\
\end{array}
$$

As this example suggests, finding solutions of even a modest 3-SAT instance can become difficult quite easily (in this case, $P$ has only one solution: $x_1 = 1, x_2 = 1, x_3 = 0, x_4 = 1, x_5 = 0, x_6 = 0$.)

\subsection{A concise review (and justification) of classical computer simulation of quantum algorithms}

For quantum computing practitioners, classical computer simulation of quantum algorithms is crucial in order to understand and to develop intuition about the behavior of quantum systems used for computational purposes, as well as to realize the approximate behavior of practical implementations of quantum algorithms. 

Early works presented by \"Omer in \cite{omer00}, Bettelli {\it et al} in \cite{bettelli03} and  Viamontes \emph{et al} in \cite{viamontes03} among others, introduced the idea of implementing quantum algorithms simulators using classical computer languages. Later and among many other interesting contributions to this field, Nyman proposed using symbolic classical computer languages for simulating quantum algorithms \cite{nyman09},  \"Omer proposed abstract semantic structures for modelling quantum algorithms in classical environments \cite{omer05}, and Altenkirch \emph{et al} proposed a quantum programming language based on classical functional programming \cite{altenkirch05}. Along with these efforts, several software platforms were developed in order to simulate quantum algorithms, being \cite{quantiki} a comprehensive list of currently available classical simulators of quantum algorithms. More recently, the availability of massively distributed computer systems like grids, clouds and GPUs has attracted the attention of researchers interested in harnessing those parallel platforms for simulating quantum algorithms, being the work produced by De Raedt \cite{deraedt07}, Caraiman \cite{caraiman10} and this paper some examples of this emerging multidisciplinary interest.

In addition to the arguments provided at the beginning of this section, another attractive application of research results on classical simulation of quantum systems is the realization of what exactly is quantum about quantum algorithms, for the following reasons: 

\begin{enumerate}
\item We need to understand exactly which properties and operations of quantum systems cannot be efficiently simulated by classical systems (see \cite{nielsenknillgotesman00} and \cite{browne07} for most interesting results related to this topic).
\item
We also need to realize whether and how exclusively quantum mechanical properties and operations can be employed for algorithm speed-up.
\end{enumerate}

An example of the importance of realizing whether truly quantum properties can be used for algorithm speed-up was provided in the field of quantum walks a few years ago. Since the publication of \cite{nayak00} it had been believed that the enhanced variance of position distribution in quantum walks was  responsible (partially at least) for quadratic speed-up of quantum walk-based  algorithms. However, arguments in favor of the plausibility of using classical physics for building experiments which replicate some interference and statistical properties of  quantum walks are given in \cite{jeong04}, \cite{knight03}, \cite{knight03b},  and \cite{knight04}, where it was shown that it is possible to develop implementations of a quantum walk on a line purely described by classical physics (wave interference of  electromagnetic fields) and still be able to reproduce the variance enhancement that characterizes a discrete quantum walk. For example, the implementation proposed in \cite{knight03b} utilizes the frequency of a light field as walker and the spatial path or the polarization state of the same light field as the coin.

\section{Reflections on natural parallel processing  and computer parallel processing}

Nature has developed very quick shortcut procedures in order to reach stable configurations (as in the case of protein folding \cite{anfinsen72}) as well as to exhaustively compute all possible configurations of a physical system (as in the case of quantum superposition and quantum parallelism \cite{bouwmeester10}). If we think of these phenomena from a computer science perspective, it is indeed our opinion that it is reasonable to hypothesize that Nature uses parallel procedures in order to quickly arrive at stable configurations, as well as to fully run natural phenomena for which an exponential or factorial amount of computer power would be needed for exhaustively computing all possible values or solutions. The question, if such a conjecture is to be further explored, is to discover \emph{how} Nature executes such parallel procedures.

A long-term goal of the authors is to find out how massively parallel computer platforms can be used for simulating parallel processes performed by Nature.  This goal includes not only running independent computations (as it would be the case the computations of all possible values of certain boolean functions using quantum parallelism) but also to find out how to simulate correlations and emergent properties of massively connected networks (e.g. \cite{scale_free_networks_review}.) This paper is a first step towards such a goal.

\section{Simulating Adiabatic Quantum Algorithms on GPUs}

\subsection{GPGPUs}
When the first Graphic Processor Units (GPUs) were designed, they were intended to support the complex mathematical operations and rendering required to create visually intensive simulations (see, for example, \cite{shirley09,devadoss11}.) As they evolved, these GPUs attracted the attention of scientists from other disciplines looking  for alternative methods to access high performance computation. This gave birth to the general purpose graphic processing units or GPGPUs. 
The Graphic Card manufacturer NVIDIA{\small$^{\copyright}$} soon became one of the most important companies creating single-chip multi-core GPUs, and they combined it with a software programming interface called CUDA which allowed  programmers to easily take advantage of parallel processing in their personal computers (\cite{birk11,nvidia08}.)  Today, NVIDIA{\small$^{\copyright}$} allows millions of users to create parallel versions of their algorithms and simulations without requiring  access to grids or clouds. It offers GPUs with up to 1024 independent cores running at 1.5GHz and their hardware can be controlled from  programs in languages like C, C++, Java, Python and many others (see, for example, \cite{sanders10}.)

In order to take full advantage of the multi-core parallelism, a program is first analized and segmented in sequential and parallel  functions. Sequential functions are preferably  run by the CPU as there is no significant processing gain in running sequential algorithms in multiple cores. Parallel code, the one without data dependencies, is consolidated in one or several kernels. Then, the programmer identifies the number of parallel execution threads required to complete the requested operation. In the case of NVIDIA{\small$^{\copyright}$} GPUs, these threads are divided into virtual blocks and grids. Threads inside a block can communicate with each other using shared memory but if two blocks of threads need to communicate, they must do so using the global memory, which is slower than the shared memory. There are physical limitations to the number of threads a block can contain and to the number of  blocks a grid can contain, bounded by the number of real cores present in the GPU. Programmers can overcome some of these limitations via developing further computer code but  the number of actual threads running at the same time can never surpass the real number of cores. 

As in any other distributed processing infrastructure, there is a need to send information from a central processing element to the distributed processing units. In the case of NVIDIA{\small$^{\copyright}$} GPUs, CUDA offers different means to send information from the CPU to the GPU including memory copy, paged memory and asynchronous communications. This process can slow down the computation and even result in worse performance than a serial approach if the design does not carefully take into account the way data is manipulated inside the GPU \cite{birk11,nvidia08}, hence the need to employ highly specialized computer programmers for this purpose.

\subsection{Mathematica and GPGPU}

The CUDA programming interface allows code from other programming languages to interact with the NVIDIA{\small$^{\copyright}$} hardware. This interaction allows the creation of higher level applications which hide the inner complexity of distributing parallel tasks to several cores to the final user. On the other hand, in the realm of quantum computing, we have been working with Mathematica{\small$^{\copyright}$} to create a high-level and high-performance simulation environment.  This high-level application has been compiled into an add-on called Quantum{\small$^{\copyright}$} (\cite{quantumitesm}), which allows end users to simulate calculations using a Dirac notation interface. Mathematica{\small$^{\copyright}$} provides ways to communicate the native user interface with code outside the package in a variety of programming languages such as C, C++ and Java \cite{cmathematica}. The combination of these two worlds led to the idea of building a bridge between the Quantum{\small$^{\copyright}$} add-on running on Mathematica{\small$^{\copyright}$}  and C++ code that could distribute processing to the GPUs.

\begin{figure}
\begin{center}
\scalebox{0.3}{\includegraphics{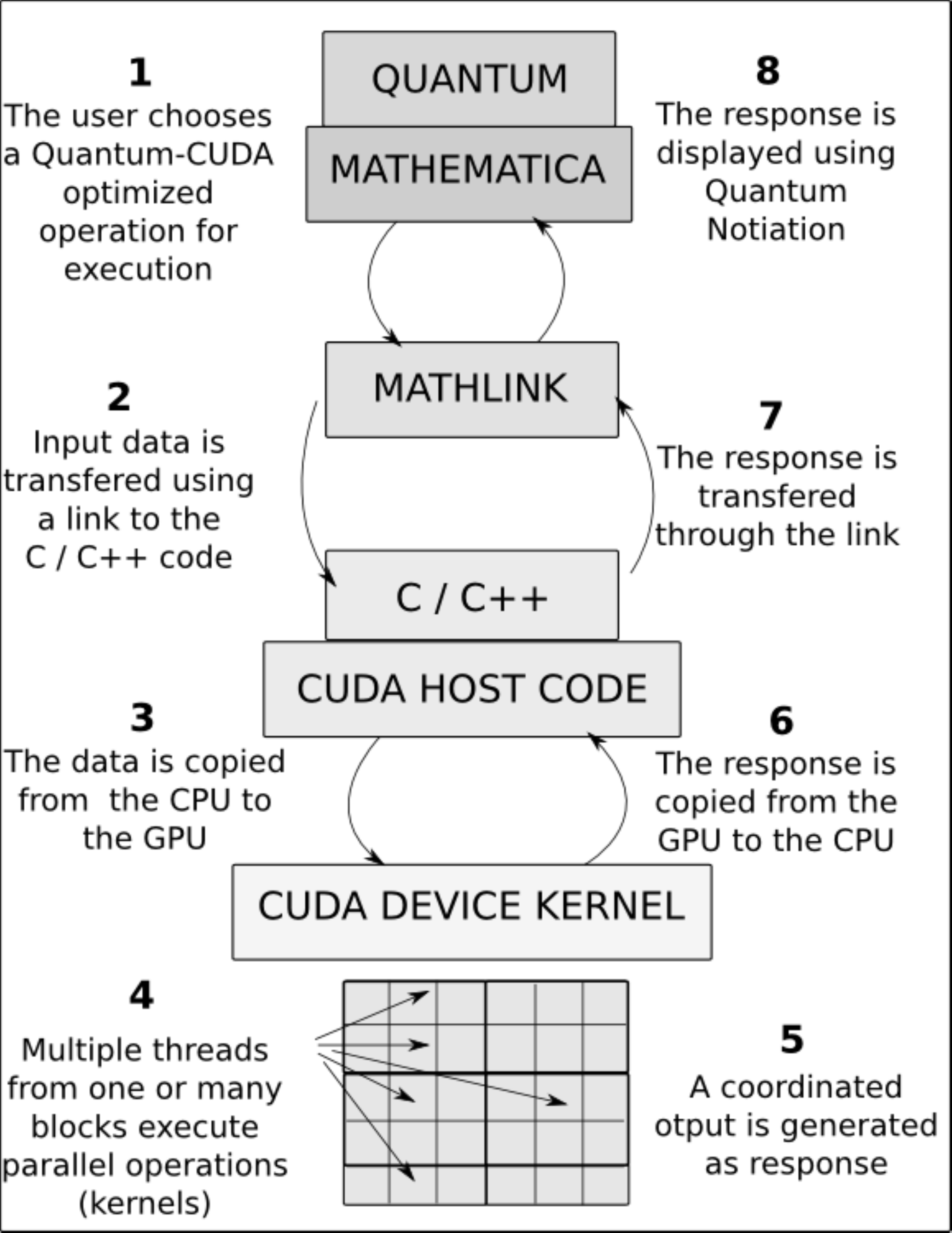}}
\end{center}
\caption{Information flow among Quantum, Mathematica, Mathlink, C/C++, CUDA, and GPUs.}
\label{flujoinfo}
\end{figure}

In our platform we have given Quantum{\small$^{\copyright}$} the ability to interact with tailored functions in CUDA to attack specific problems or to communicate with the specialized linear algebra libraries CUBLAS and CULA. This has enabled us to enhance monolithic simulations or atomic operations within a complex simulation. So far we have worked with Mathematica{\small$^{\copyright}$} 7 and Mathlink{\small$^{\copyright}$}: we have created the data structures using the high-end interface of Quantum{\small$^{\copyright}$}, have then sent this information using Mathlink{\small$^{\copyright}$} to a C++ code that deploys blocks of threads in the GPU to satisfy corresponding requests.  A result is built using information from every thread and then sent back to Quantum{\small$^{\copyright}$} using Mathlink{\small$^{\copyright}$}. This process gives Quantum{\small$^{\copyright}$} 
users the ability to use their desktop or laptop computers as high performance computation infrastructure to simulate quantum algorithms and quantum processes in  a very user friendly manner. Mathematica{\small$^{\copyright}$} 8 now integrates a native way to interact with CUDA which allows the deployment of kernels directly and without using Mathlink{\small$^{\copyright}$}, thus we expect our kernels to run faster and in a more integrated way than in current Mathematica{\small$^{\copyright}$} 7.  We present in Fig. (\ref{flujoinfo}) a visualization of data flow among Quantum, Mathematica, Mathlink, C/C++ code, CUDA, and GPU hardware.

\begin{figure}
\begin{center}
\scalebox{0.3}{\includegraphics{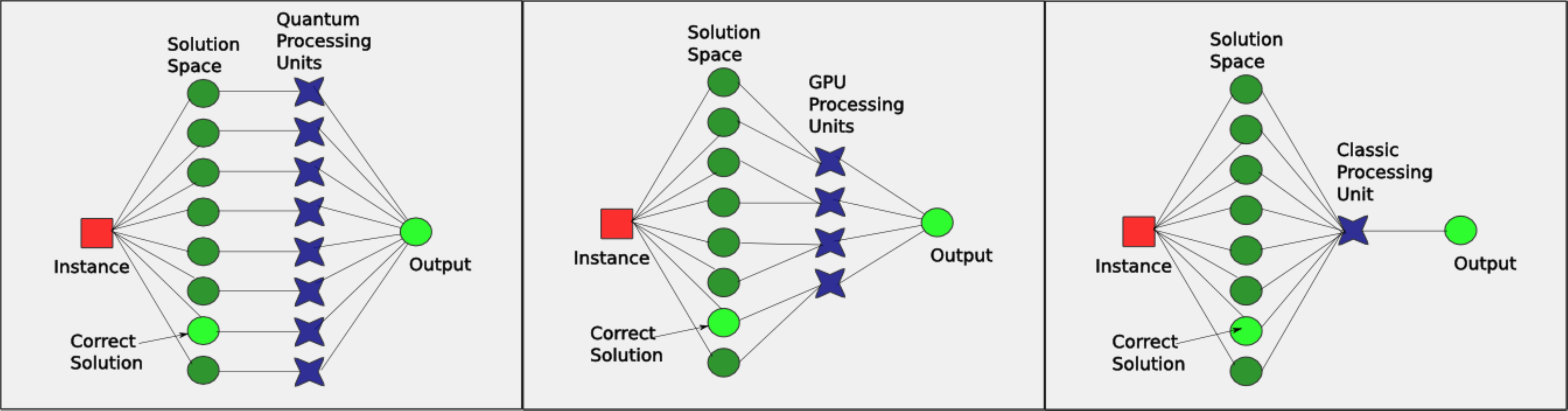}}
\end{center}
\caption{Quantum, multi-core, and serial computational approaches.}
\label{tres_segmentos}
\end{figure}

In order to stress differences among parallel and serial simulation of quantum processes,  we show in Fig. (\ref{tres_segmentos}) these three different computational approaches to solve an instance of a problem. On the left hand side segment of Fig. (\ref{tres_segmentos}) of we see an algorithm which uses quantum processing units to solve the problem at hand: taking advantage of quantum parallelism, we use only one processing unit for all the solution space, so the computational load per processing unit is low. In the central portion of Fig. (\ref{tres_segmentos}) we see a Multi-core Multi-thread GPU based approach to solve the problem. Here, the number of cores is limited, but the possible solutions are distributed among the available cores so the computational load is higher than in a quantum approach but lower than a serial approach. In the right segment of Fig. (\ref{tres_segmentos}) we see a classical serial implementation to solve the problem, in which the computational load per processing unit increases because all the possible solutions must be tested in only one core.

One interesting feature of this project is the ease with which new kernels can be written. It is very common that one of the biggest obstacles for scientists using parallel infrastructure is to be able to transform their serial algorithms into corresponding parallel versions. Sometimes this process is not even suitable for the application and results in worse performance than the stand-alone approach. Nevertheless, taking quantum algorithms and deploying them into parallel structures is easier because they are already engineered to exploit the quantum parallelism.

\subsection{Results}

We have tested our system by building a software platform for simulating an adiabatic quantum algorithm for solving hard instances of the 3SAT problem \cite{perdomo11}.  The adiabatic quantum algorithm we have simulated consists of the design of a time-dependent Hamiltonian which can be separated into three parts. The first part, the initial Hamiltonian, encodes the ground state of the system that should be easy to prepare. The second part, the driving Hamiltonian, is in charge of taking the system from an initial state to the final state. The third part, the final Hamiltonian, is created from an energy function which will give every possible state an energy  level proportional to the number of unsatisfied clauses. The energy function depends on the instance and is constituted by a sum of  smaller energy functions, one for each clause. The ground state of this final Hamiltonian encodes the solution to the problem.

With the purpose of exhibiting the advantages of using GPUs instead of CPUs for quantum algorithm simulation, we have firstly run a simulation of the above-mentioned algorithm over a CPU. Then, we built a specific CUDA kernel to enhance its performance. The idea of our parallel 
implementation consists of simulating quantum parallelism with multi-core parallelism. We took the energy function over which the above-mentioned Hamiltonian is built and turned it into a kernel. This way, we can create multiple processing threads and each will evaluate one combination of variables and assign it an energy level using the function. 

We tested our simulation environment with instances of the 3SAT problem using a ratio from number of clauses to number of variables about $4.2$. The tests were run using a PC with Intel Core 2 Duo processor @ 2.66GHz, 8GB of RAM memory running with Windows Vista and an NVIDIA Geforce GTX 8800 video card of 512MB of video memory and 128 parallel cores. The simulation environment currently runs on Mathematica 7.

\begin{figure}
\begin{center}
\scalebox{0.3}{\includegraphics{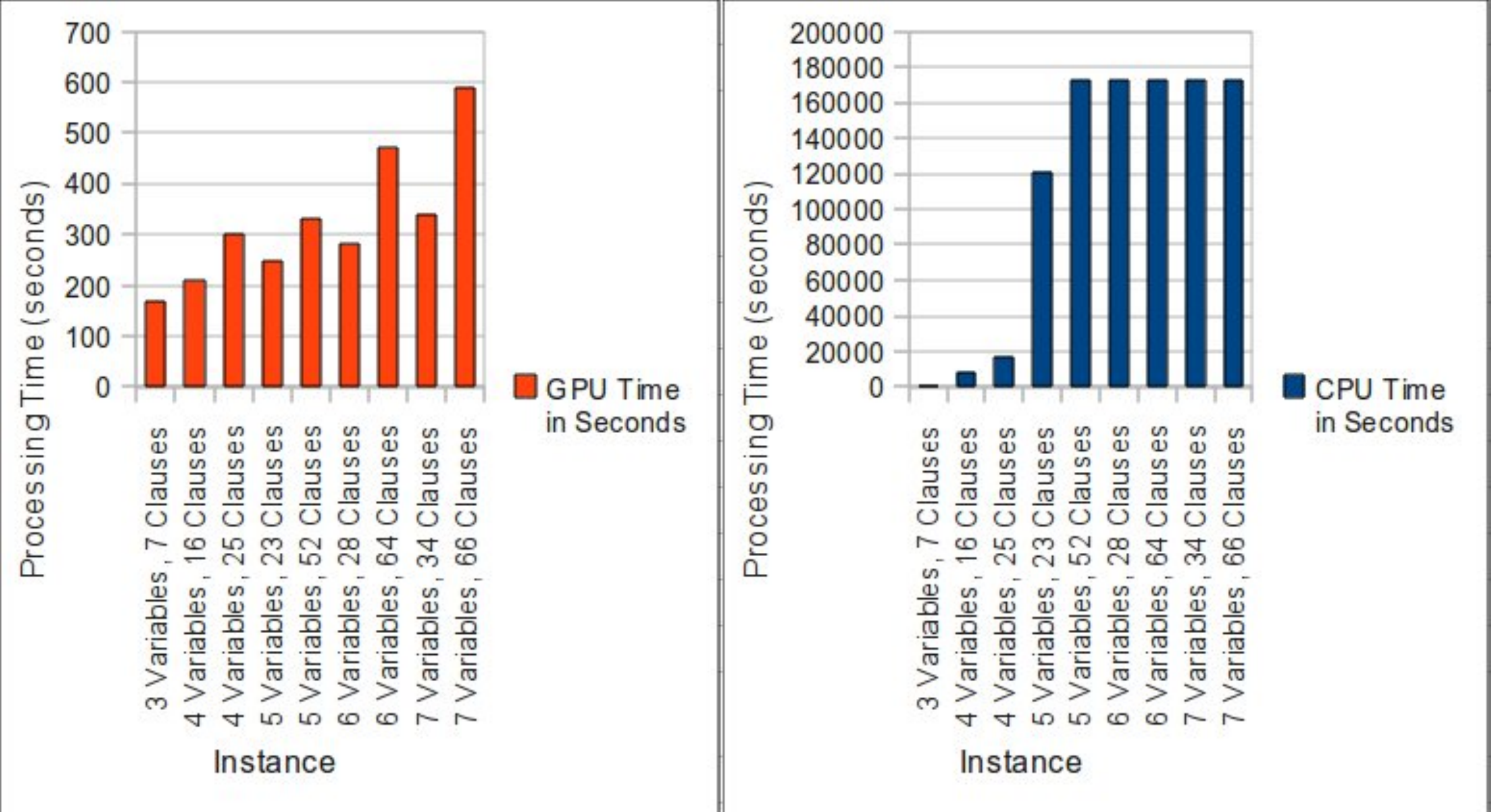}}
\end{center}
\caption{CPU and GPU execution times for different instances of the 3SAT problem}
\label{uno}
\end{figure}

\begin{figure}
\begin{center}
\scalebox{0.3}{\includegraphics{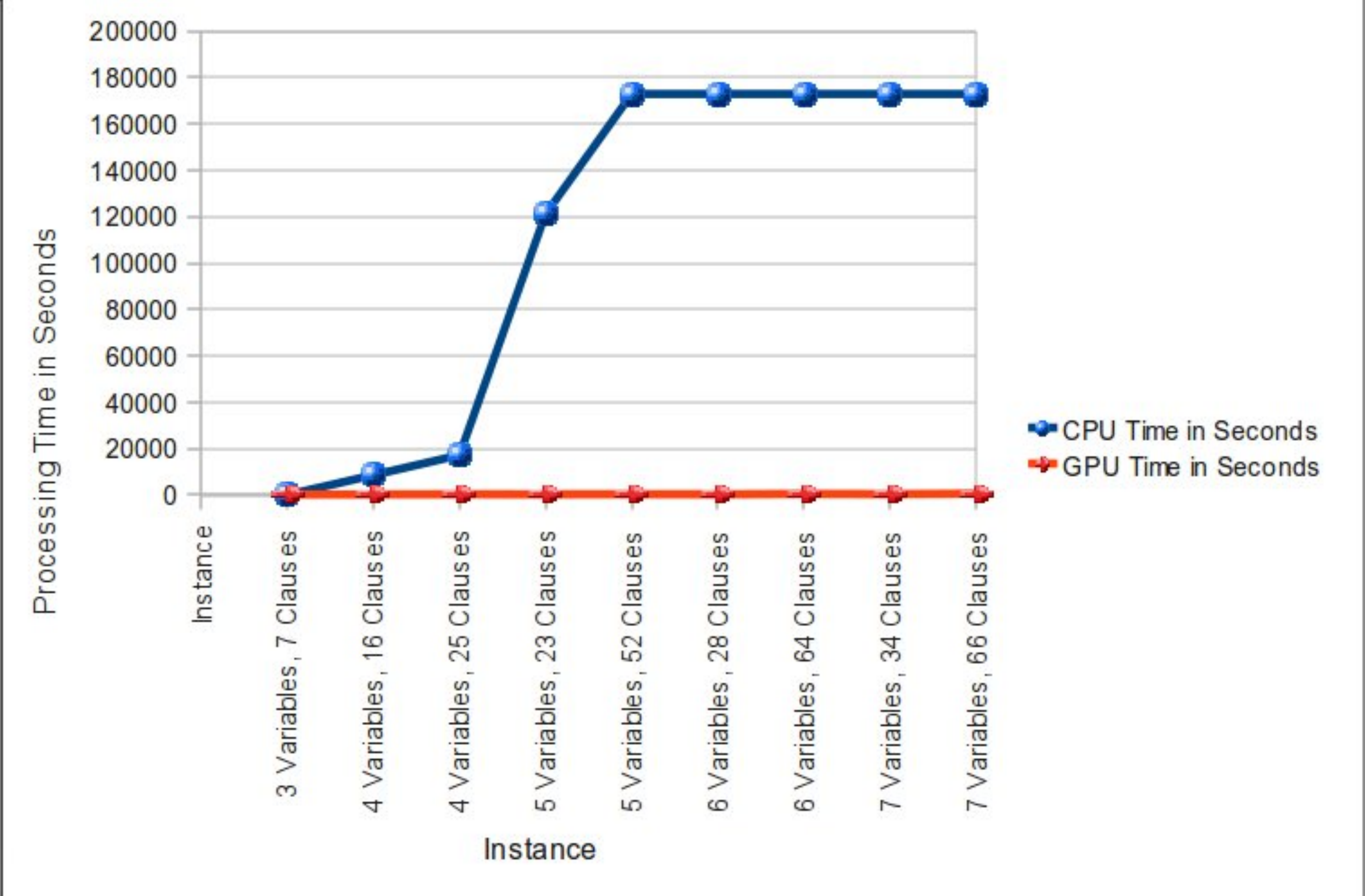}}
\end{center}
\caption{Comparison between CPU and GPU execution times for different instances of the 3SAT problem}
\label{dos}
\end{figure}

\begin{figure}
\begin{center}
\scalebox{0.3}{\includegraphics{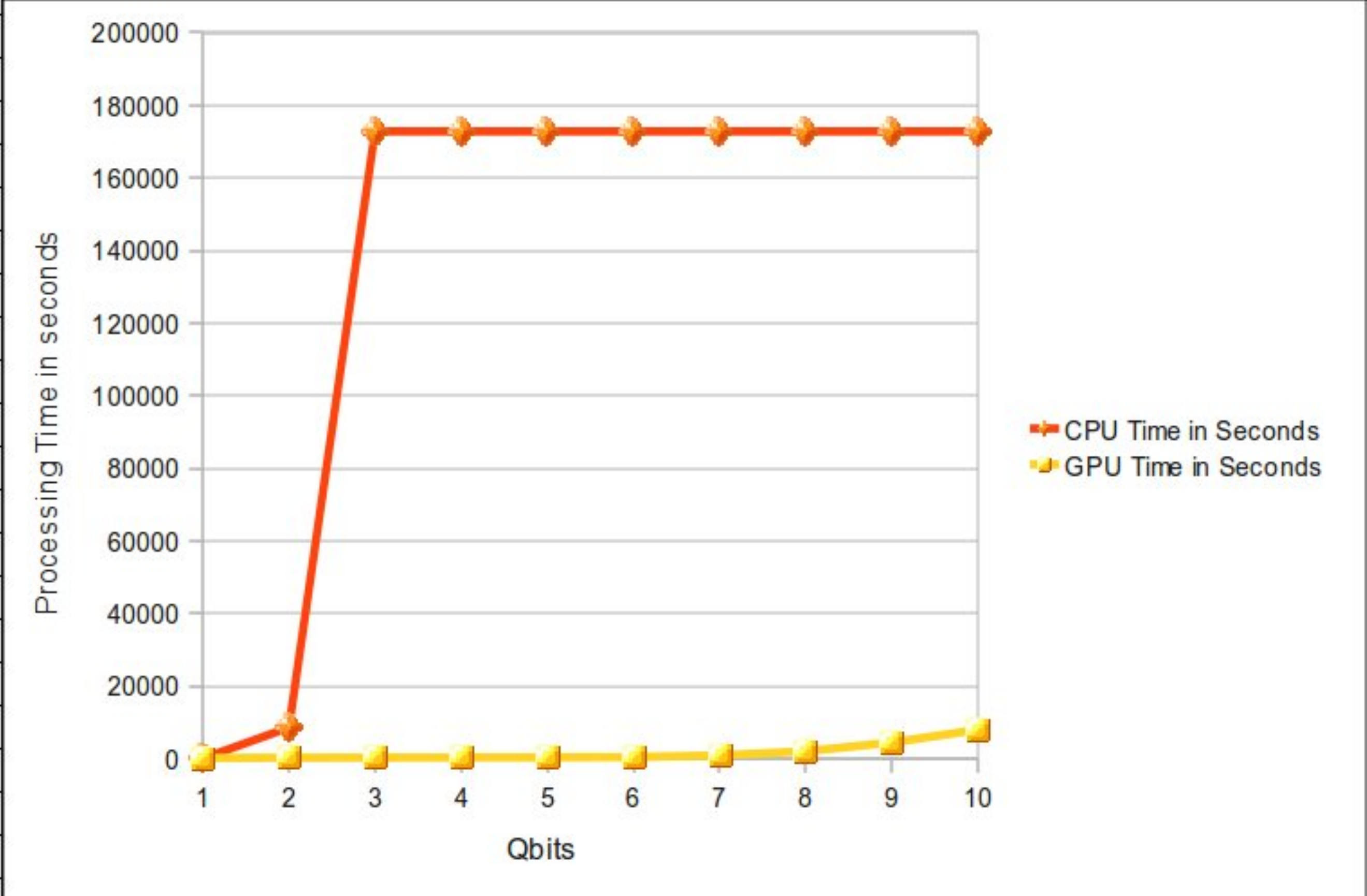}}
\end{center}
\caption{Comparison between CPU and GPU simulation times for different number of qubits}
\label{tres}
\end{figure}

In Fig. (\ref{uno}) we present the results obtained in processing time for different instances of the 3SAT problem running on the CPU and the GPU. In Fig. (\ref{dos}) we show a comparison between the results obtained with both devices. As it can be seen in Figs. (\ref{uno},\ref{dos}), the processing time used by the CPU increases exponentially while the time in the GPU increases on an slower ratio and
scales according to the GPU occupancy factor, i.e. the number of actual parallel cores required to fulfill the processing needs of each instance. In Fig. (\ref{tres}) we can see the processing time used to simulate instances of the 3SAT for several qubits. These results were limited to the instances that could be simulated within a 2.5 days processing time frame.

Based on our results, we observe that the number of qubits simulated using our GPU tools easily double the ones 
simulated on a CPU using our setup. These results are mainly due to the combination of two characteristics in our simulation: 
firstly, we aid the simulation tasks with the power of multi-core GPU processing with kernels designed to take advantage of the special
memory, thread management and synchronization capabilities of NVIDIA cards. Secondly, we simulate quantum parallelism directly with
classical multi-core parallelism, which allows us to exploit the GPU occupancy factor to the maximum on every run.

Even when the number of possible variable combinations surpasses the available number of parallel threads in the GPU, we can still get 
an excellent performance enhancement. We use shared memory inside each processing block to enhance the access 
time to data within the kernel. We also write the result to global data concurrently, in separated memory blocks, to enhance the data 
throughput.


\section{Conclusions}

In this paper we have presented a GPU-based symbolic and parallel platform for clasically simulating quantum algorithms. Our simulation environment is based on a high-level user interface developed on  Mathematica{\small$^{\copyright}$} which is connected to C++ code in order to make our platform capable of communicating with a Graphics Processing Unit. The main contribution of this work is the creation of a simulation environment enhanced with parallel processing which can be used on personal  computers and creates a direct comparison between quantum parallelism and classic multi-core parallelism. 

In order to properly introduce the behavior of our proposal we have simulated a quantum adiabatic algorithm designed for solving hard instances of the 3-SAT problem. Based on our results, we observe that the number of qubits that can be simulated using our GPU tools doubles the ones simulated on a CPU efficiently using our setup. These results are possible due to the combination of two characteristics in our simulation:  firstly, we aid the simulation tasks with the power of multi-core GPU processing with kernels designed to take advantage of the special memory, thread management and synchronization capabilities of NVIDIA cards, and secondly, we simulate quantum parallelism directly with
classical multi-core parallelism, which allows us to maximally exploit the GPU occupancy factor on every run.  Additionally, we have presented a review of currently available classical simulators of quantum systems together with some justifications, based on our willingness to further understand processing properties of Nature, for devoting resources and efforts to building more powerful simulators.


\section{Acknowledgments}
All authors  acknowledge the financial support of ITESM-CEM. S.E. Venegas-Andraca is also thankful for the financial support of CONACyT and SNI (41594).




\end{document}